\documentclass[journal]{IEEEtran}
\usepackage{color}
\usepackage{leftidx}
\usepackage{cite}
\usepackage{microtype}
\usepackage{cleveref}
\usepackage{graphicx,amssymb,amstext,amsmath,cases,subfigure}
\usepackage[ruled,vlined,lined,ruled,linesnumbered]{algorithm2e}
\begin{document}

\title{Multi-antenna Enabled Cluster-based Cooperation in Wireless Powered Communication Networks}

\author{Lina Yuan, Suzhi Bi, \IEEEmembership{Member,~IEEE,} Shengli Zhang, \IEEEmembership{Senior Member,~IEEE,} Xiaohui Lin, and Hui Wang\\
\thanks{The authors are with the College of Information Engineering, Shenzhen University, Shenzhen,  Guangdong, China 518060. E-mail:~2156130110@email.szu.edu.cn, \{bsz, zsl, xhlin, wanghsz\}@szu.edu.cn.}}

\maketitle

\begin{abstract}
In this paper, we consider a wireless powered communication network (WPCN) consisting of a multi-antenna hybrid access point (HAP) that transfers wireless energy to and receives sensing data from a cluster of low-power wireless devices (WDs). To enhance the throughput performance of some far-away WDs, we allow one of the WDs to act as the cluster head (CH) that helps forward the messages of the other cluster members (CMs). However, the performance of the proposed cluster-based cooperation is fundamentally limited by the high energy consumption of the CH, who needs to transmit all the WDs' messages including its own. To tackle this issue, we exploit the capability of multi-antenna energy beamforming (EB) at the HAP, which can focus more transferred power to the CH to balance its energy consumption in assisting the other WDs. Specifically, we first derive the throughput performance of each individual WD under the proposed scheme. Then, we jointly optimize the EB design, the transmit time allocation among the HAP and the WDs, and the transmit power allocation of the CH to maximize the minimum data rate achievable among all the WDs (the max-min throughput) for improved throughput fairness among the WDs. An efficient optimal algorithm is proposed to solve the joint optimization problem. Moreover, we simulate under practical network setups and show that the proposed multi-antenna enabled cluster-based cooperation can effectively improve the throughput fairness of WPCN.
\end{abstract}

\begin{IEEEkeywords}
Wireless sensor networks, wireless powered communication, resource allocation, user fairness.
\end{IEEEkeywords}

\section{Introduction}
\IEEEPARstart{T}{he} performance of modern communication networks is largely constrained by the limited battery life of wireless devices (WDs). Once the energy is depleted, a WD needs manual replacement/recharging of its battery, which can result in frequent interruption to normal device operation and severe communication performance degradation. Alternatively, the recent development of wireless energy transfer (WET) technology enables a novel networking paradigm named wireless powered communications network (WPCN) \cite{2015:Bi,2015:Lu,2016:Bi}, where the information transmissions of WDs can be continuously and remotely powered by the microwave energy transmitted by dedicated energy nodes. The use of WET can effectively reduce the battery replacement/recharging cost and also improve the communication quality with reduced energy outages. With its potential to tackle the critical energy constraints, we can expect that WET will be an important building block in future wireless communication networks.

There are extensive studies on implementing WPCN in low-power applications, such as wireless sensor network (WSN) and radio frequency identity (RFID) networks \cite{2012:Xie}, to prolong the network operating lifetime or increase the data rate \cite{2016:Bi1,2016:Bi2,2014:Huang}. In a WPCN, the energy node and the information access point (that receives information from WDs) can either be separately located or co-located as a hybrid access point (HAP) \cite{2016:Bi1}. While the former scheme enjoys more degree of freedom in device placement, the latter can save network deployment cost and is easier for the HAP to centrally coordinate the energy and information transmissions. In this paper, we focus on studying a WPCN using a HAP for both energy provision and information access.

The throughput performance of a multi-user WPCN coordinated by a HAP is first studied in \cite{2014:Ju}, which proposes a harvest-then-transmit protocol that applies the HAP to first broadcasts radio frequency (RF) energy to all WDs in the downlink, and then the WDs transmit their individual information with time-division-multiple-access (TDMA) to HAP using their harvested energy in the uplink. It is also revealed in \cite{2014:Ju} that such design will lead to severe user unfairness problem, namely the ``doubly near-far" problem, due to distance-dependent power loss. In particular, some devices' data rates can be two orders of magnitude smaller than the others, which directly decreases the sensing accuracy of a WPCN. One effective method to improve the throughput fairness is through user cooperation, where close-to-HAP users help forward the messages of far-away users \cite{2014:Ju1,2015:Chen,2017:Zhong}. Interestingly, \cite{2014:Ju1} shows that by helping the far-away user in a two-user WPCN, the close-to-HAP user can also improve its data rate, resulting a win-win situation. Further, the two-user cooperation is later studied when the two cooperating users form a distributed virtual antenna array for information transmission in \cite{2017:Zhong} and extended to a general multiple user cooperation scenario in \cite{2015:Chen}.

The above studies on the throughput performance of WPCN mostly consider using a single-antenna HAP and focus on optimizing the transmit time allocation to improve the throughput performance. The single-antenna HAP, however, suffers from very low energy transfer efficiency due to the fast signal power attenuation of omnidirectional energy transmission. Instead, when the HAP is equipped with multiple antennas, it can apply energy beamforming (EB) technique \cite{2013:Zhang} to focus the transferred energy to desired directions to enhance the energy transfer efficiency to specific devices, and thus the data rates of energy-harvesting devices. The optimal EB design has been studied in several practical setups, e.g., training sequence design \cite{2015:Zeng}, hardware feedback complexity constraints \cite{2014:Xu}, and under per-antenna transmit power constraint \cite{2017:Rezaei}. Besides, the multiple antennas can also improve the communication performance by leveraging spatial diversity or multiplexing gains in the uplink.

A number of recent works have considered the design of WPCN when a multi-antenna HAP is applied. For instance, \cite{2014:Liu} first studies the optimal energy and information beamforming design and uplink/downlink transmit time allocation, and showed the use of multiple antenna can significantly improve the throughput performance compared to its single-antenna counterpart in \cite{2014:Ju}. The throughput optimization is then studied in \cite{2015:Yang} when the HAP has a large number of antennas (i.e., massive MIMO). Nonetheless, the doubly-near far problem in WPCN still exists regardless of the number of antennas at the HAP. Therefore, cooperation methods are also widely adopted when multi-antenna HAP is concerned. For instance, \cite{2017:Liang} considers using a fixed single-antenna relay to forward the message of an energy-harvesting user to a multi-antenna HAP, and studies the optimal beamforming design and transmit time allocation. \cite{2017:Xiong} proposes a group collaboration where two communication groups cooperate with each other under the coordination of a multi-antenna HAP.

\begin{figure}
  \centering
  \includegraphics[width=0.45\textwidth]{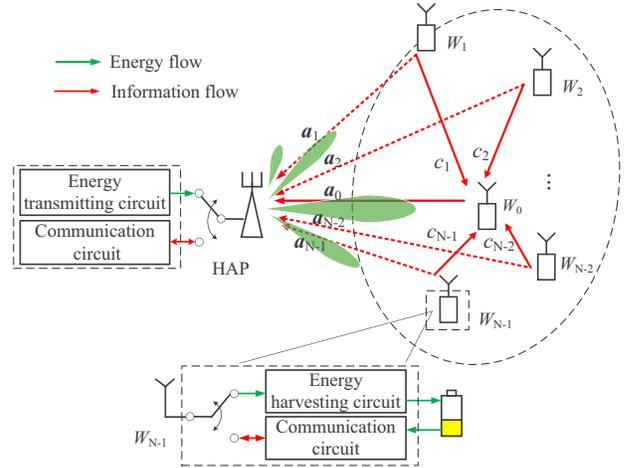}\\
  \caption{A schematic of the considered cluster-based cooperation in WPCN, where W$_0$ is the cluster head and the rest $(N-1)$ WDs are cluster member.}
  \label{101}
\end{figure}

In this paper, we consider a cluster-based user cooperation in a WPCN as shown in Fig.~\ref{101}, where a multi-antenna HAP applies WET to power a cluster of remote WDs and receives their data transmissions. This may correspond to a practical scenario in WSNs, where a mobile HAP pauses in its route to power a cluster of densely deployed sensors monitoring a particular area. Like a conventional WSN, we designate one of the WDs as the cluster head (CH) to forward the information transmission of the other cluster members (CMs) to the HAP. Intuitively, the throughput performance of some far-away WDs can be improved thanks to the cooperation. However, like cluster-based cooperation in conventional WSN (e.g., \cite{2009:Chen}), the CH inevitably suffers from high energy consumption as it needs to transmit all the users' messages including its own. For a cluster with a large number of WDs, the CH's limited battery will become the performance bottleneck of the network. To solve this energy imbalance problem, we propose to exploit the capability of multi-antenna energy beamforming at the HAP, where the HAP can focus more transferred power to the CH to balance the energy consumption in assisting other WDs. The detailed contributions of this paper are as follows.
\begin{itemize}
  \item We propose a cluster-based cooperation method in WPCN, where a WD is designated as the CH to forward the information transmissions of the other sensors. To address the high energy consumption of the CH in conventional cluster-based cooperation scheme, we apply EB technique at the multi-antenna HAP to balance the different energy consumption rates of the WDs.
  \item With the proposed cooperation method, we formulate a joint optimization problem of EB design, the transmit time allocation among the HAP and WDs, and the transmit power allocation of the CH, to maximize the minimum data rate achievable among all the WDs (i.e., the max-min throughput) for improved user fairness. An efficient optimal solution algorithm is proposed to solve the non-convex optimization problem.
  \item We perform numerical analysis to study the impact of different system setups to the performance of the proposed method. By comparing with other benchmark schemes, we show that the proposed cooperation can effectively improve the throughput performance. Besides, the proposed cooperation method is most effective when the WD that is closest to the cluster center is selected as the CH, the WDs are closely located with strong intra-cluster channels, and the number of cooperating WDs is moderate to support efficient cooperations.
\end{itemize}

The rest of the paper is organized as follows. We introduce the system model and propose the cluster-based cooperation method in Section II. We analyze the per-WD throughput performance in Section III. In Section IV, we formulate the maxi-min throughput optimization problem and propose optimal solution algorithm. In Section V, we evaluate the performance of the proposed cooperation using simulations. Finally, the paper is concluded in Section VI.

\section{System Model}
\subsection{Channel Model}
As shown in Fig.1, we consider a WPCN consisting of a HAP and $N$ WDs. The HAP is equipped with $M$ antennas ($M<<N$ in practice), while each WD is equipped with one single antenna. Specifically, the HAP broadcasts wireless energy to and receives wireless information transmission (WIT) from the WDs. The HAP has stable power supply and each WD has a rechargeable battery to store the harvested wireless energy from the HAP. The HAP and all the WDs operate over the same frequency band, where a time division duplexing (TDD) circuit \cite{2013:Zhou} is implemented at both the HAP and the WDs to separate the energy and information transmissions.

In this paper, one of the WDs is selected as the CH that helps relay the WIT of the other CMs. The impact of CH selection method to the system performance will be discussed in Section V. Without loss of generality, the CH is indexed as W$_0$, and the CMs are indexed as W$_1$, $\cdots$, W$_{N-1}$. All the channels are assumed to be independent and reciprocal and follow quasi-static flat-fading, such that all the channels coefficients remain constant during each block transmission time, denoted by $T$, but can vary from in different blocks. The channel coefficient vector between the HAP and W$_{i}$ is denoted by $\mathbf{a}_{i} \in\mathcal{C}^{M\times1}$,  where $\mathbf{a}_{i} \sim \mathcal{CN} (\mathbf{0}, \sigma_{i}^2 \mathbf{I})$ and $\sigma_i^2$ denotes the average channel gain, $i=0,1,\cdots, N-1$. Besides, the channel coefficient between the $j$-th CM and the CH is denoted by $c_j\sim \mathcal{CN} ( 0, \delta_{j}^2)$, $j=1,\cdots, N-1$. Here, we use $h_{i} \triangleq |\mathbf{a}_{i}|^2$ and $g_i\triangleq |c_i|^2$ to denote the corresponding channel gains, where $|\cdot|$ denotes the 2-norm operator.

\begin{figure}
  \centering
  \includegraphics[width=0.45\textwidth]{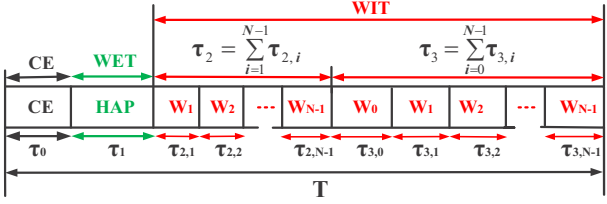}\\
  \caption{The proposed cluster-based cooperation protocol in WPCN.}
  \label{102}
\end{figure}

\subsection{Cluster-based Cooperation Protocol}
The operation of the proposed cluster-based cooperation in a transmission time block is illustrated in Fig.~\ref{102}. At the beginning of a transmission block, channel estimation (CE) is performed within a fixed duration $\tau_{0}$. During the CE stage, the WDs take turns to broadcast their pilot signals, so that HAP has the knowledge of $\mathbf{a}_i$, $i=0,1,\cdots, N-1$, and the CH knows $c_i$, $i=1, \cdots, N-1$, respectively. Then, the CH sends its estimation of $c_i$'s to the HAP, such that the HAP has the full knowledge of CSI in the network.

After the CE stage, the system operates in three phases. In the first phase with time duration $\tau_{1}$, the HAP broadcasts wireless energy with fixed transmit power $P$. In the next two phases with $T-\tau_{0}-\tau_{1}$ amount of time, the $N$ WDs transmit their independent information to the HAP using their individually harvested energy. Specifically, the $(N-1)$ CMs first transmit in turn to the CH, where the $i$-th CM transmits for $\tau_{2,i}$ amount of time, $i=1, \cdots, N-1$. In the third phase, the CH transmits the decoded messages of the $(N-1)$ CMs along with its own message to the HAP. The time taken to transmit the $i$-th WD's message is denoted as $\tau_{3,i}$, $i=0,1,\cdots, N-1$, Evidently, the time allocations satisfy the following inequality
\begin{equation}
\label{1}
    \tau_0+\tau_1+\sum^{N-1}_{i=1}\tau_{2,i}+\sum^{N-1}_{i=0}\tau_{3,i} \leq T.\\
\end{equation}
Notice that $\tau_0$ is a known parameter. Without loss of generality, we assume $T=1$ throughout this paper. Based on the knowledge of global CSI, the HAP can calculate the optimal time allocation and then broadcast to all the WDs such that they can keep their time-switching circuit synchronized for either energy and information transmission. Notice that, besides the transmission in the third phase, the HAP can also overhear each CM's message in the second phase, although not dedicated to it, which can be used to improve the overall transmission rate compared to decoding the message in the third phase alone. In the next section, we derive the throughput performance of the proposed cooperation protocol and formulate the max-min throughput optimization problem.

\section{Per-WD Throughput Analysis}
In this section, we derive the throughput of each WD achieved by the proposed cluster-based cooperation protocol. The results will be used in the next section to optimize the throughput fairness of the WPCN.
\subsection{Phase I: Energy Transfer}
We notice that the CH needs to transmit $N$ messages in total, which would consume significantly more energy than the other CMs, making CH the performance bottleneck of the network. To balance the energy consumed and harvested for each WD, the HAP adopts EB to deliver different power to WDs located in different directions. Specifically, in the first phase of time $\tau_1$, the HAP transmits $\mathbf{w}(t) \in C^{M\times 1}$ random energy signals on the $M$ antennas, where the transmit power of HAP is constrained by
\begin{equation}
\label{2}
E\left[|\mathbf{w}(t)|^2\right] = \text{tr}\left(E\left\{\mathbf{w}(t)\mathbf{w}(t)^H\right\}\right) \triangleq \text{tr}(\mathbf{Q}) \leq P.
\end{equation}
where $\text{tr}(\cdot)$ denotes the trace of a matrix, $(\cdot)^H$ denotes the complex conjugate operator, and $\mathbf{Q}\succeq \mathbf{0}$ is the beamforming matrix. Then, the received energy signal by the $i$-th WD is
\begin{equation}
\label{3}
y_i^{(1)}(t) = \mathbf{a}_i^T\mathbf{w}(t) + n^{(1)}_i(t), i = 0, \cdots, N-1,
\end{equation}
where $n_i^{(1)}(t)$ denotes the receiver noise power. With the noise power neglected, the amount of energy harvested by the WDs can be expressed as [7]
\begin{equation}
\label{4}
E_{i}=\eta \tau_1 E\left[|y_i^{(1)}(t)|^2\right] = \eta \tau_1 \cdot \text{tr}(\mathbf{A}_i\mathbf{Q}).
\end{equation}
Here, $\mathbf{A}_i \triangleq \mathbf{a}_i\mathbf{a}_i^H$ and $\eta\in(0,1]$ denotes the energy harvesting efficiency, which is assumed equal for all the WDs.

\subsection{Phase II: Intra-cluster Transmissions}
We assume that the CMs exhaust the harvested energy to transmit to the CH during the second stage. Then, the transmit power of the $i$-th CM is
\begin{equation}
\label{5}
P_{2, i}=\frac {E_{i}}{\tau_{2,i}} =\eta \frac{\tau_1}{\tau_{2,i}}\text{tr}(\mathbf{A}_i\mathbf{Q}),\  i=1,\cdots,N-1.
\end{equation}
Let $s_{i}^{(2)}(t)$ denote the baseband signal of the $i$-th WD transmitted in the second phase with $E[|s_{i}^{(2)}(t)|^2]=1$, the received signal at the CH is expressed as
\begin{equation}
\label{6}
\begin{aligned}
   &y_{0,i}^{(2)}(t)= c_i \sqrt{P_{2, i}}s_{i}^{(2)}(t)+n_{i}^{(2)}(t),
\end{aligned}
\end{equation}
where $n_{i}^{(2)}(t)$ denotes the receiver noise with power $E\left[|n_{i}^{(2)}(t)|^2\right]=N_0$. Then, the CH can decode the $i$-th CM's message at a rate given by
\begin{equation}
\label{7}
\begin{aligned}
R_{i}^{(2)}&=\tau_{2,i} \log_{2}\left(1 + \frac{g_i P_{2,i}}{N_0}\right),i=1,\cdots,N-1.
\end{aligned}
\end{equation}
Meanwhile, the HAP can also overhear the transmission of the CMs, such that it receives
\begin{equation}
\label{8}
\mathbf{y}_{H,i}^{(2)}(t) = \mathbf{a}_i \sqrt{P_{2, i}}s_{i}^{(2)}(t)+ \mathbf{n}_{H,i}^{(2)}(t).
\end{equation}
during the $i$-th CM's transmission, where $i=1,\cdots,N-1$, and $\mathbf{n}_{H,i}^{(2)}(t) \sim \mathcal{CN}(\mathbf{0},N_0\mathbf{I})$. For simplicity, we neglect the energy consumption on decoding and consider only data transmission consuming the harvested energy. However, the proposed method can be easily extended to the case with non-zero decoding energy consumption by including a constant circuit power term.

\subsection{Phase III: Cluster-to-HAP Transmission}
After decoding the CMs' messages, the CH transmits the $(N-1)$ CMs' messages along with its own message one by one to the HAP. Let $s_{0}^{(3)}(t)$ denote CH's baseband signal and $s_{i}^{(3)}(t)$ denote the re-encoded baseband signal of the $i$-th CM transmitted in the third phase. Besides, we assume $E[|s_i^{(3)}(t)|^2]=1$, $i=0,\cdots,N-1$. Let $P_{3,i}$ denote the power used to transmit the $i$-th WD's message. Then, the received signal of the $i$-th WD's message at the HAP is
\begin{equation}
\label{9}
\mathbf{y}_{i}^{(3)}(t) = \mathbf{a}_0 \sqrt{P_{3,i}}s_{i}^{(3)}(t) + \mathbf{n}_{i}^{(3)}(t),\ i=0,1,\cdots, N-1.
\end{equation}
The total energy consumed by CH is upper bounded by its harvested energy $E_0$, i.e.,
\begin{equation}
\label{10}
\sum^{N-1}_{i=0}\tau_{3,i} P_{3,i} \leq \eta \tau_1 \text{tr}(\mathbf{A}_0 \mathbf{Q}).
\end{equation}

We assume that the HAP uses maximal ratio combining (MRC) to maximize the receive signal-to-noise power ratio (SNR), where the combiner output SNR of the $i$-th WD is
\begin{equation}
\label{11}
\begin{aligned}
\gamma^{(3)}_{i} = \frac{|\mathbf{a}_0|^2 P_{3,i}}{N_0}=\frac{h_0 P_{3,i}}{N_0}, \ i=0,\cdots, N-1.
\end{aligned}
\end{equation}
Denote the time allocation as $\pmb{\tau}=[\tau_1, \tau_{2,1}, \cdots, \tau_{2,N-1}, \tau_{3,0}, \\ \tau_{3,1}, \cdots, \tau_{3,N-1}]'$, and the transmit power as $\mathbf{P}=[P_{3,0}, P_{3,1}, \cdots, P_{3,N-1}]'$. Then, the data rate of the CH at the HAP is
\begin{equation}
\label{12}
\begin{aligned}
R_{0}(\pmb{\tau}, \pmb{P})&= \tau_{3,0} \log_{2}\left(1 + \frac{h_0 P_{3,0}}{N_0}\right).
\end{aligned}
\end{equation}

For each CM's message, however, is received in both the second and third phases. In this case, the HAP can jointly decode each CM's message across two phases at a rate given by \cite{2014:Ju}
\begin{equation}
\label{13}
R_{i}(\pmb{\tau}, \pmb{P}, \mathbf{Q}) = \min\left\{R_{i}^{(2)}(\pmb{\tau}, \mathbf{Q}), V_{i}^{(2)}(\pmb{\tau}, \mathbf{Q}) + V_{i}^{(3)}(\pmb{\tau}, \pmb{P})\right\}.
\end{equation}
where $i=1, \cdots,N-1$, and $R_{i}^{(2)}(\pmb{\tau}, \mathbf{Q})$ is given in (\ref{7}).
$V_{i}^{(2)}(\pmb{\tau}, \mathbf{Q})$ denotes the information that can be extracted by the HAP from the received signal in (\ref{6}) (in the second phase) using an optimal MRC receiver, which is given by
\begin{equation}
\label{14}
\begin{aligned}
V_{i}^{(2)}(\pmb{\tau}, \mathbf{Q})&=\tau_{2,i} \log_{2}\left(1 + \eta \frac{\tau_1}{\tau_{2,i}} \frac{h_i\text{tr}(\mathbf{A}_i\mathbf{Q})}{N_0}\right).
\end{aligned}
\end{equation}
$V_{i}^{(3)}(\pmb{\tau}, \pmb{P}, \mathbf{Q})$ denotes the achievable rates of the transmissions from CH to the HAP, which are given by
\begin{equation}
\label{15}
\begin{aligned}
V_{i}^{(3)}(\pmb{\tau}, \pmb{P})&=\tau_{3,i} \log_{2}\left(1+\frac{h_0 P_{3,i}}{N_0}\right).
\end{aligned}
\end{equation}

An important performance metric of a WPCN is the max-min throughput, defined as
\begin{equation}
S = \min_{0 \leq i \leq N-1} R_i,
\end{equation}
i.e., the minimum achievable per-WD throughput, which reflects the throughput fairness among the WDs. The max-min throughput has important practical implication. For instance, the max-min throughput in a WSN reflects the accuracy of data reported by the ``bottleneck" sensor, which can directly affect the overall sensing accuracy of the network. In the next section, we formulate the max-min throughput optimization problem and solve it optimally. In fact, our proposed method in this paper can also be extended to maximize (weighted) sum throughput of the WDs, which is omitted for brevity.

\section{Max-min Throughput Optimization}

\subsection{Problem Formulation}
In this section, we are interested in maximizing the minimum (max-min) throughput of all WDs in each block, by jointly optimizing the energy beamforming $\mathbf{Q}$, the time allocation $\pmb {\tau}$, and the transmit power allocation $\pmb P$, i.e.,
\begin{equation}
\label{16}
   \begin{aligned}
(P1):\; &\max_{\mathbf{\pmb {\tau, P}}, \mathbf{Q}}& &S= \min_{0 \leq i \leq N-1} R_i(\pmb {\tau, P}, \mathbf{Q})\\
        &\text{s. t.}& & (1) \;\rm{and} \; (10), \\
                 & & &\tau_{1} \geq 0, \; \tau_{2,i} \geq 0, \;i=1,\cdots, N-1,\\
                 & & & \tau_{3,i} \geq 0,\;P_{3,i}\geq 0,\;i=0,1,\cdots, N-1,\\
                 & & & \text{tr}(\mathbf{Q}) \leq P, \;\mathbf{Q} \succeq \mathbf{0}, \; \pmb {\tau} \geq \mathbf{0}.\\
    \end{aligned}
\end{equation}
By introducing a variable $\overline S$, problem (\ref{16}) can be equivalently transformed into its epigraphic form,
\begin{equation}
\label{17}
   \begin{aligned}
(P2):\quad &\max_{\pmb {\tau}, \pmb{P}, \mathbf{Q}, \overline S} & &  \overline S\\
    &\text{s. t.}  & & (1) \;\rm{and} \; (10), \\
                   & & & R_{0}(\pmb{\tau}, \pmb{P}) \geq \overline S,\\
                   & & &V_{i}^{(2)}(\pmb{\tau}, \mathbf{Q}) + V_{i}^{(3)}(\pmb{\tau}, \pmb{P}) \geq \overline S,\\
                   & & &R_{i}^{(2)}(\pmb{\tau}, \mathbf{Q}) \geq \overline S, i=1,\cdots, N-1,\\
                   & & & \text{tr}(\mathbf{Q}) \leq P, \;\mathbf{Q} \succeq \mathbf{0}, \; \pmb {\tau} \geq \mathbf{0}.\\
    \end{aligned}
\end{equation}
Due to joint design of user cooperation and energy beamforming, both the data rates in intra-cluster communication (i.e., $R_{i}^{(2)}(\pmb{\tau}, \mathbf{Q})$ and $V_{i}^{(2)}(\pmb{\tau}, \mathbf{Q}))$ and cluster-to-HAP communication (i.e., $R_{0}(\pmb{\tau}, \pmb{P})$ and $V_{i}^{(3)}(\pmb{\tau}, \pmb{P})$) are not concave functions. Besides, the LHS of (\ref{10}) is also not a convex function. Therefore, (P2) is a non-convex problem in its current form, which lacks of efficient optimal algorithm. In the next subsection, we propose an algorithm to solve (P2) optimally.

\subsection{Optimal Algorithm to (P2)}
We first define $\mathbf{W} \triangleq \tau_1\mathbf{Q} \succeq \mathbf{0}$. With the sum transmit power constraint in (\ref{2}), we have
\begin{equation}
\text{tr}\left(\mathbf{W}\right) = \text{tr}\left(\tau_1\mathbf{Q}\right) \leq \tau_1 P.
\end{equation}
Accordingly, we change the variables as
\begin{equation}
z_i \triangleq \tau_1\text{tr}\left(\mathbf{A}_i\mathbf{Q}\right) = \text{tr}\left(\mathbf{A}_i\mathbf{W}\right),
\end{equation}
for $i=0,\cdots, N-1$. Thus, $R_{i}^{(2)}(\pmb{\tau}, \mathbf{Q})$ and $V_{i}^{(2)}(\pmb{\tau}, \mathbf{Q})$ in (\ref{7}) and (\ref{14}) can be re-expressed as functions of $\pmb {\tau}$ and $\pmb z=\left[z_{1}, \cdots, z_{N-1}\right]'$,
\begin{equation}
\label{18}
R_{i}^{(2)}(\pmb{\tau},\pmb {z})=\tau_{2,i} \log_{2}\left(1+ \overline{\rho}_i \frac { z_i}{\tau_{2,i}}\right),
\end{equation}
\begin{equation}
\label{19}
V_{i}^{(2)}(\pmb{\tau},\pmb {z})=\tau_{2,i} \log_{2}\left(1+ \rho_i \frac {z_i}{\tau_{2,i}}\right),
\end{equation}
where $i=1,\cdots,N-1$, and $\overline {\rho}_i\triangleq\eta \frac{g_i}{N_0}$ and $\rho_i\triangleq\eta \frac{h_i}{N_0}$ are parameters.

Subsequently, we define $\theta_{3,i}\triangleq\frac{\tau_{3,i} P_{3,i}}{\eta}$, $i=0,1,\cdots,N-1$, then $R_{0}(\pmb{\tau}, \pmb{P})$ and $V_{i}^{(3)}(\pmb{\tau}, \pmb{P})$ in (\ref{12}) and (\ref{15}) can be reformulated as  functions of $\pmb {\tau}$ and $\pmb {\theta}=\left[\theta_{3,0}, \cdots, \theta_{3,N-1}\right]'$, i.e.,
\begin{equation}
\label{20}
\begin{aligned}
R_{0}(\pmb {\tau, \theta})= \tau_{3,0} \log_{2}\left(1 + \rho_0 \frac { \theta_{3,0}}{\tau_{3,0}}\right),
\end{aligned}
\end{equation}
\begin{equation}
\label{21}
\begin{aligned}
V_{i}^{(3)}(\pmb {\tau, \theta})=\tau_{3,i} \log_{2}\left(1+\rho_0 \frac { \theta_{3,i}}{\tau_{3,i}}\right).
\end{aligned}
\end{equation}
where $i=1,\cdots,N-1$, and $\rho_0\triangleq\eta \frac{h_0}{N_0}$. Thus, the power constraint given in (\ref{10}) can be re-expressed as
\begin{equation}
\label{22}
\begin{aligned}
\sum^{N-1}_{i=0}\theta_{3,i}\leq z_0.
\end{aligned}
\end{equation}

Accordingly, problem (\ref{17}) can be transformed into the following equivalent problem.
\begin{equation*}
\label{23}
   \begin{aligned}
(P3):\quad &\max_{\pmb {\tau}, \pmb{\theta}, \pmb z,  \overline S, \mathbf{W}} & &  \overline S\\
    &\text{s. t.}  & & R_{0}(\pmb{\tau, \theta}) \geq \overline S,\\
                   & & &V_{i}^{(2)}(\pmb{\tau},\pmb{z}) + V_{i}^{(3)}(\pmb{\tau, \theta}) \geq \overline S,\\
                   & & &R_{i}^{(2)}(\pmb{\tau},\pmb{z}) \geq \overline S,\ i=1,\cdots, N-1,\\
                   & & &\tau_0+\tau_1+\sum^{N-1}_{i=1}\tau_{2,i}+\sum^{N-1}_{i=0}\tau_{3,i} \leq 1,\\
                   & & &z_i=\text{tr}(\mathbf{A}_i\mathbf{W}),\ i=0,1,\cdots, N-1,\\
                   & & &\sum^{N-1}_{i=0}\theta_{3,i}\leq z_0,\ \pmb {\tau} \geq \mathbf{0},\\
                   & & & \text{tr}(\mathbf{W}) \leq \tau_1 P,\ \mathbf{W}\succeq \mathbf{0}.
    \end{aligned}
\end{equation*}
Before solving (P3), we have the following Lemma 1.

$\underline{Lemma} \; \emph {1:}$ When $x>0$ and $y>0$, $z=x\log_2(1+y/x)$ is jointly concave in ($x$, $y$).

\emph {Proof:} The Hessian of $z(x, y)$ is
\begin{equation}
\label{24}
\bigtriangledown^2 z (x, y)=
\frac{1}{\ln 2 (x+y)^2}\left[
  \begin{array}{cc}
  -\frac{y^2}{x} & y\\
    y   & -x\\
\end{array}
\right]
\end{equation}
When $x, y>0$, for any arbitrary vector $\mathbf{d} = (d_1,d_2)'$, we have
\begin{equation}
\mathbf{d}' \cdot \bigtriangledown^2 z \cdot \mathbf{d} = -\frac{\left(\frac{d_1y}{\sqrt{x}}-d_2\sqrt{x}\right)^2}{\ln 2 (x+y)^2} \leq 0.
\end{equation}
Therefore, $\bigtriangledown^2 z$ is a negative semi-definite matrix, which completes the proof.  $\hfill\blacksquare$.

From Lemma 1, we can see that both $R^{(2)}_{i}$'s in (\ref{18}) and $V_{i}^{(2)}$'s in (\ref{19}) are concave functions in $(\pmb{\tau},\pmb{z})'$. Besides, $R_{0}$ in (\ref{20}) and $V_{i}^{(3)}$'s in (\ref{21}) are also concave functions in $(\pmb {\pmb{\tau}, \theta})'$. Therefore, the first three sets of constraints in (P3) are convex constraints. Meanwhile, the rest of the constraints are affine. Accordingly, it follows that the objective and all the constraints of (P3) are convex, therefore (P3) is a convex optimization problem, which can be efficiently solved by off-the-shelf optimization algorithms, e.g., interior point method \cite{2004:Boyd}. Let's denote the optimal solution to (P3) as $\left\{\pmb {\tau}^*, \pmb{\theta}^*, \pmb z^*,  \overline S^*, \mathbf{W}^*\right\}$. Then, the optimal solution $\pmb {\tau}^*$ of (P1) is the same as that in (P3). The optimal $\mathbf{Q}^*$ and $\mathbf{P}^*$ of (P1) can be restored by letting $\mathbf{Q}^* = \mathbf{W}^*/\tau_1^*$ and $P_{3,i}^* = \eta\theta_{3,i}^*/\tau_{3,i}^*,\ i = 0,\cdots,N-1.$

\subsection{Benchmark Methods}
For performance comparison, we consider two representative benchmark methods. For simplicity, we assume that the time spent on CE ($\tau_0$) is equal for all the schemes.
\subsubsection{Cluster-based cooperation w/o EB}
The only difference from the proposed cooperation method is that the HAP does not apply EB and instead transmitting wireless energy isotropically to the WDs during the WET phase. In this case, the optimal time allocation $\boldsymbol{\tau}^*$ and transmit power allocation $\mathbf{P}^*$ can be obtained by fixing $\mathbf{Q}^*=\frac{P}{M}\mathbf{I}$ in (P1), where $\mathbf{I}$ denotes an identity matrix.

\subsubsection{Independent transmission with EB}
In this case, all the WDs transmit independently to the HAP following the harvest-then-transmit protocol in \cite{2014:Ju}. Specifically, the HAP first uses EB to performs WET for $\tau_1'$ amount of time for the WDs to harvest. Then, the WDs take turns to transmit their messages to the HAP, where each WD's transmission takes $\tau_{2,i}'$ ($i=0,1,\cdots, N-1$) amount of time. Meanwhile, the HAP uses MRC to decode the message of each user.\footnote{Spatial multiplexing is not used at the HAP as the number of WDs is often much larger than the number of antennas at the HAP. Otherwise, either strong interference or high computational complexity will be induced when the WDs transmit to the HAP simultaneously.} Then, the data rate of the $i$-th user is denoted by
\begin{equation}
\label{28}
R_{i}'(\pmb{\tau}', \mathbf{Q'}) = \tau_{2,i}'  \log_{2}\left(1 + \gamma_{i}'\right), \ i=0,\cdots,N-1,
\end{equation}
where
\begin{equation}
\label{27}
\gamma_{i}' = \frac{\eta \tau_1' h_i \text{tr}(\mathbf{A}_i\mathbf{Q'})}{ N_0\tau_{2,i}'}
\end{equation}
denotes the output SNR, $\mathbf{Q'}$ denotes the beamforming matrix, and $\pmb{\tau}' \triangleq[\tau_1', \tau_{2,0}', \cdots, \tau_{2,N-1}']'$. Then, the max-min throughput can be obtained by solving the following problem
\begin{equation}
\label{29}
   \begin{aligned}
 &\max_{\mathbf{\pmb {\tau}'}, \mathbf{Q}'}& &\min_{i=0,\cdots, N-1} R_{i}'(\pmb{\tau}', \mathbf{Q'})\\
        &\text{s. t.}& & \tau_0+\tau_1'+\sum^{N-1}_{i=0}\tau_{2,i}' \leq 1, \\
                 & & & \tau_{1}' \geq 0,\;\tau_{2,i}'\geq 0, \ i=0,\cdots, N-1,\\
                 & & & \text{tr}(\mathbf{Q'}) \leq P,\; \;\mathbf{Q}' \succeq \mathbf{0}. \\
    \end{aligned}
\end{equation}
The optimal solution to the above problem can be similarly obtained as (P3), where the details are omitted for brevity.

  \begin{figure}
\centering
\subfigure[Cooperation with EB.]{
\label{figa} 
\includegraphics[width=0.45\textwidth]{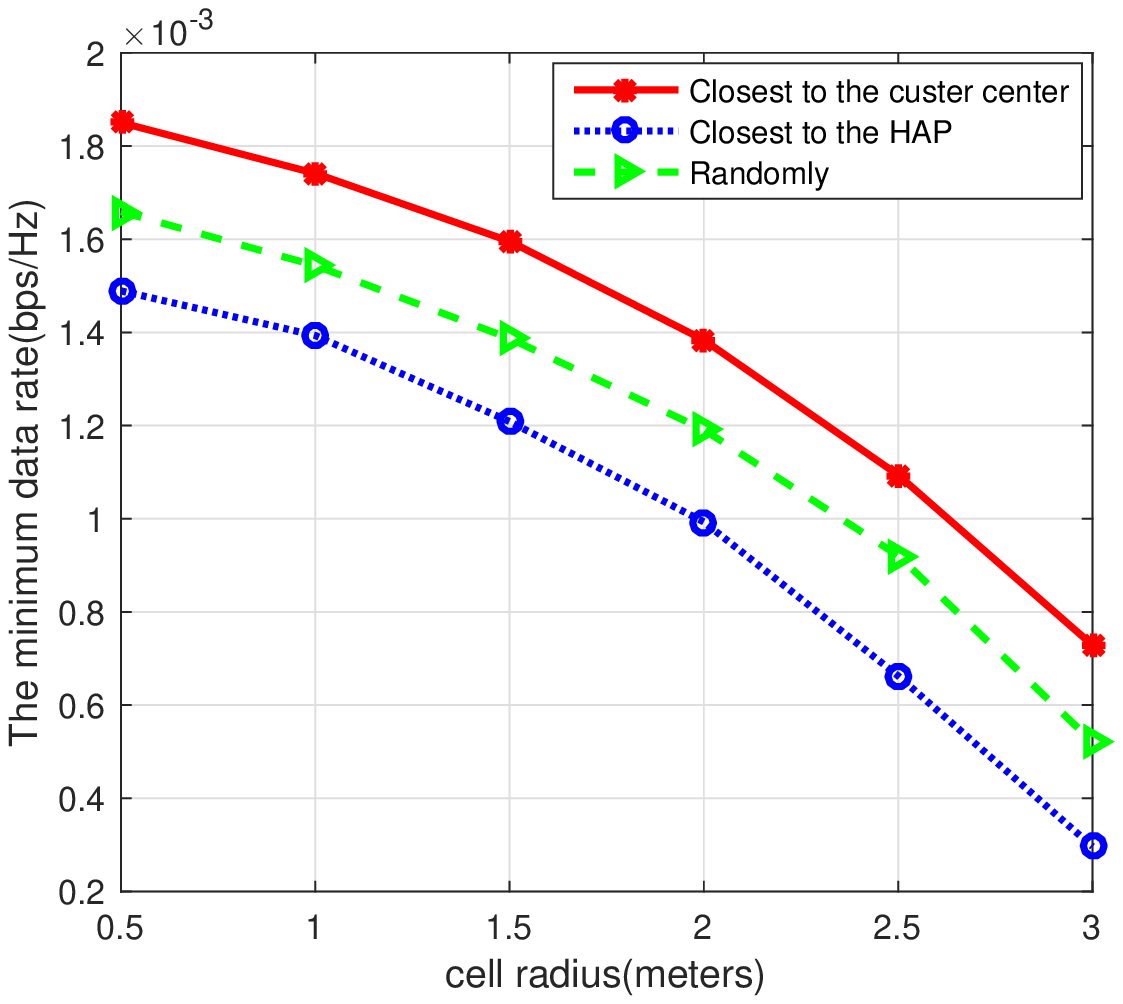}}
\hspace{1in}
\subfigure[Cooperation without EB.]{
\label{fig:subfig:b} 
\includegraphics[width=0.45\textwidth]{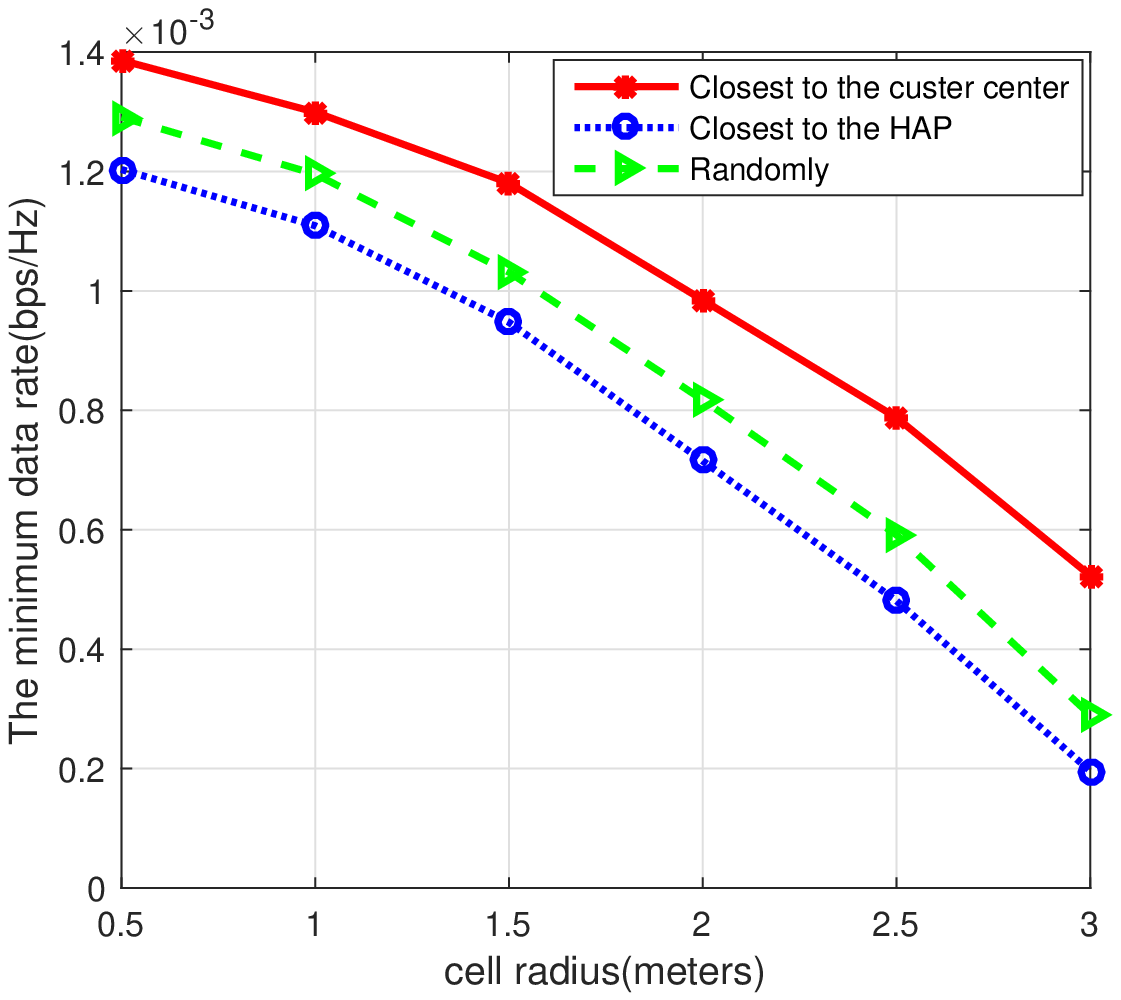}}
\caption{The impact of cluster head selection to the max-min throughput with $d=6$ meters. The figure above adopts EB technique at the HAP and the below does not.}
\label{103} 
\end{figure}

\section{Simulation Results}
In this section, we evaluate the performance of the proposed cooperation method. In all simulations, we use the Powercast TX91501-3W transmitter as the energy transmitter at the HAP with transmit power $P=3$ watts and P2110 Powerharvester as the energy receiver at each WD with $\eta= 0.51$ energy harvesting efficiency.\footnote{Please see the detailed product specifications on the website of Powercast Co. (http://www.powercastco.com).} Without loss of generality, it is assumed that the number of antennas at HAP is $M=5$ and the noise power $N_{0}$ is $10^{-10}$ $W$ in the considered bandwidth for all receivers. The mean channel gain between any two nodes, either the HAP or a WD, follows a path-loss model. For instance, let $d_{H,i}$ denote the distance between the HAP and the $i$-th WD, then the average channel gain $\delta_i^2 = G_A(\frac{3\times 10^8}{4\pi d_{H,i}f_c})^{\alpha}$, where $G_A$ denotes the antenna gain, $\alpha$ denotes the path-loss factor and $f_{c}$ denotes the carrier frequency. Unless otherwise stated, we assume $G_A=2$, $\alpha =3$, and $f_{c}=915$ $MHz$. Besides, $15$ WDs are uniformly distributed within a circle with radius equal to $r$ meters, and the circle's center is $d$ meters away from the HAP. Each point in the figures is an average of $20$ independent WD placements.

 \begin{figure}
\centering
\subfigure[Cooperation with EB.]{
\label{figa} 
\includegraphics[width=0.45\textwidth]{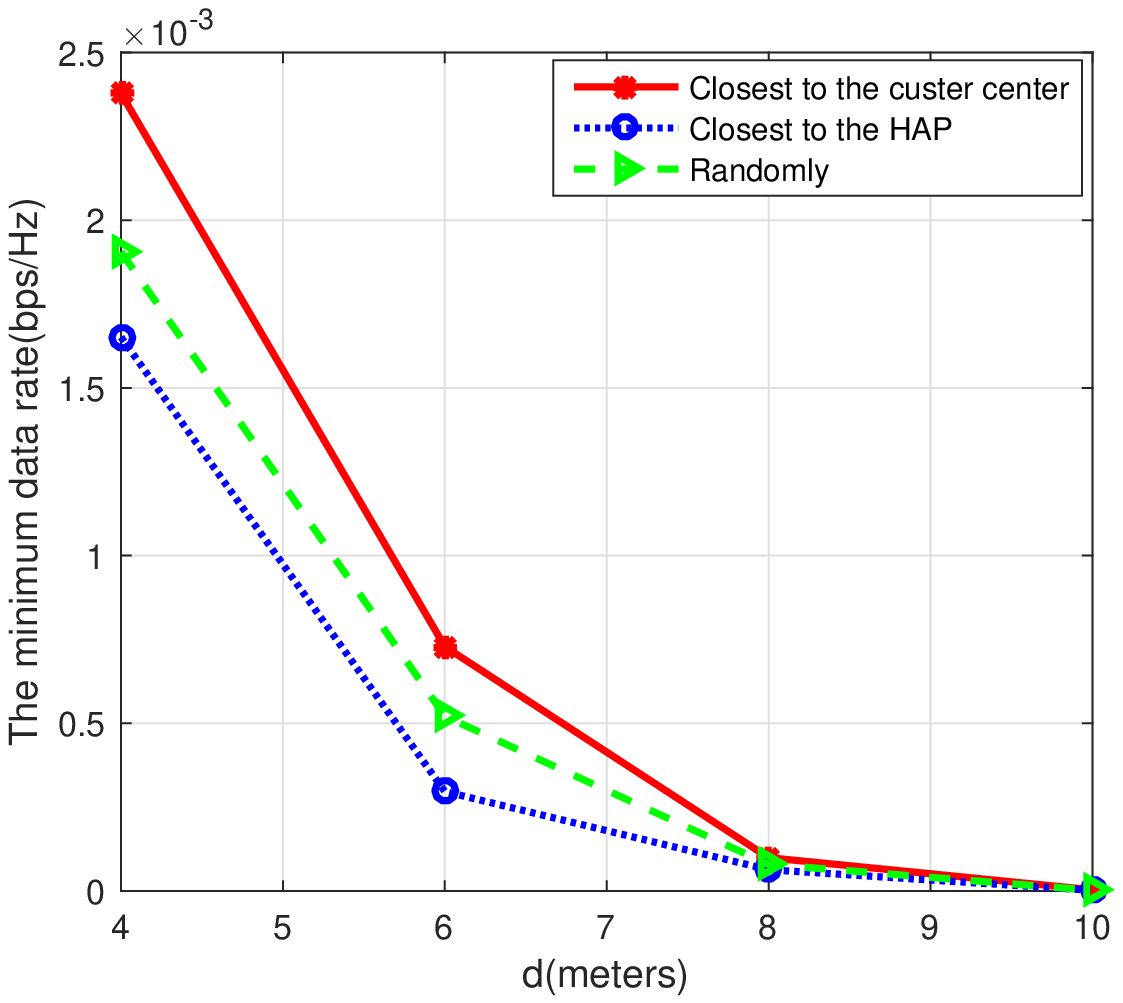}}
\hspace{1in}
\subfigure[Cooperation without EB.]{
\label{fig:subfig:b} 
\includegraphics[width=0.45\textwidth]{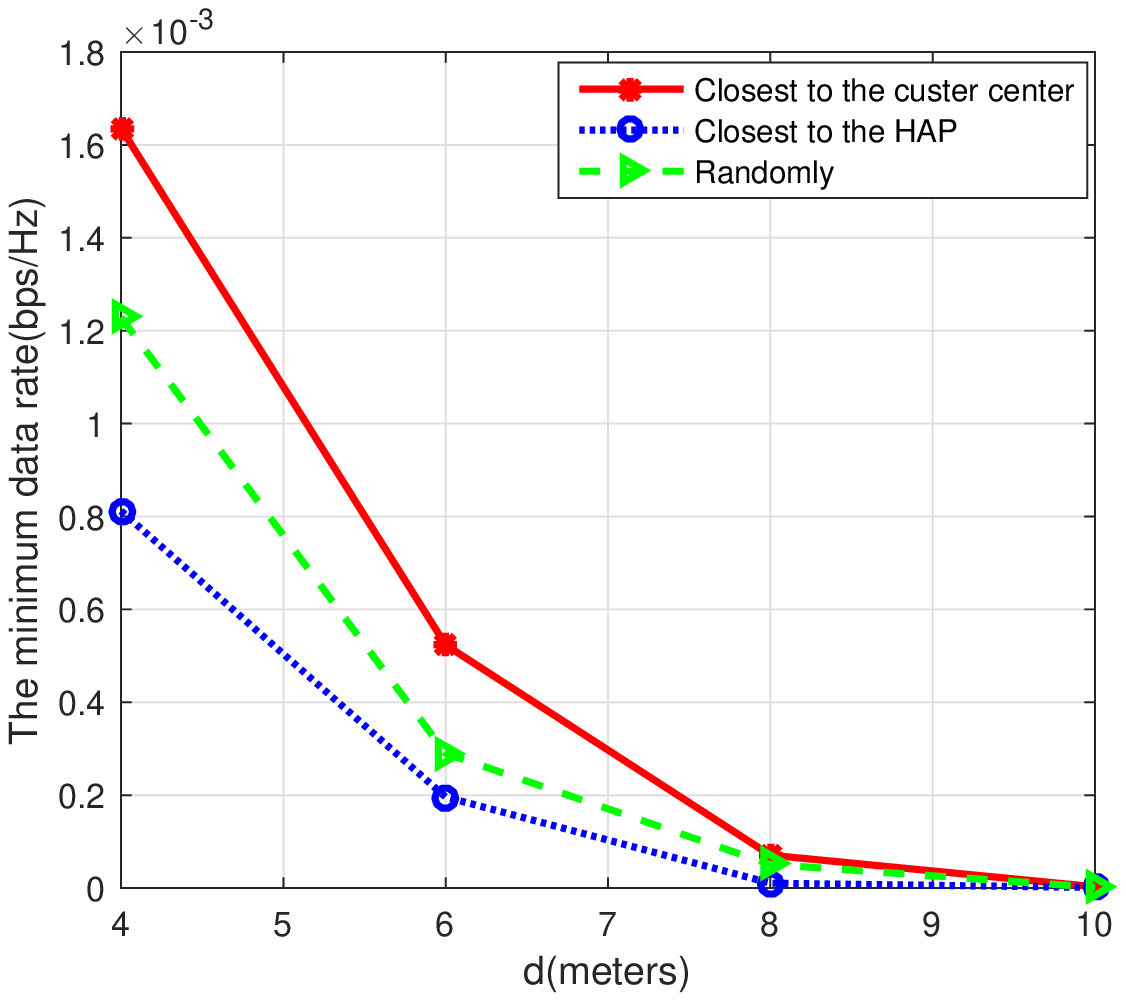}}
\caption{The impact of cluster head selection to the max-min throughput with $r=3$ meters. The figure above adopts EB technique at the HAP and the below does not.}
\label{104} 
\end{figure}

In Fig.~\ref{103}, we investigate the impact of cluster head selection method on the throughput performance. Specifically, we consider three CH selection methods: selecting the WD that is closest to the cluster center,\footnote{The location of cluster center can be obtained by taking the average of the location coordinates of the $N$ WDs} closest to the HAP, or randomly \footnote{The performance is an average of $5$ random CH selections for each WD placement.} from the WDs. Specifically, we fix the distance $d=6$ meters and change the radius of the cluster $r$, and consider two different methods with EB either adopted (the proposed cooperation method) or not (cooperation without EB in Section IV.B) at the HAP. As expected, the data rates of the three CH selection methods decrease as the cell radius increases, because the intra-cluster communication links become weaker when the distances between the CMs and the CH increase. Meanwhile, regardless of EB is used or not at the HAP, selecting the WD closest to the cluster center achieves the best performance. Interestingly, we can also see that selecting the WD closest to the HAP performs even worse than selecting a random WD as the CH. This is because, on average, the largest distance between the CMs and the CH is larger for the former scheme than the latter. Similar result is also observed in Fig.~\ref{104} when we fix the cell radius $r=3$ meters and vary the distance $d$. Both Fig.~\ref{103} and Fig.~\ref{104} show that to achieve high throughput fairness, \emph{efficient intra-cluster cooperation} is required such that the distance disparity between the CMs and the CH should be minimized, e.g., by selecting the WD closest to the cluster center. Therefore, we designate the WD \emph{closest to the cluster center as the CH} when cluster-based cooperation is considered in the following simulations.

  \begin{figure}
\centering
\subfigure[Max-min throughput.]{
\label{figa} 
\includegraphics[width=0.45\textwidth]{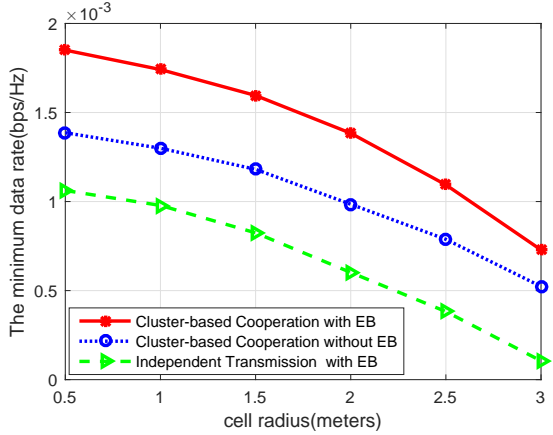}}
\hspace{1in}
\subfigure[Sum throughput.]{
\label{fig:subfig:b} 
\includegraphics[width=0.45\textwidth]{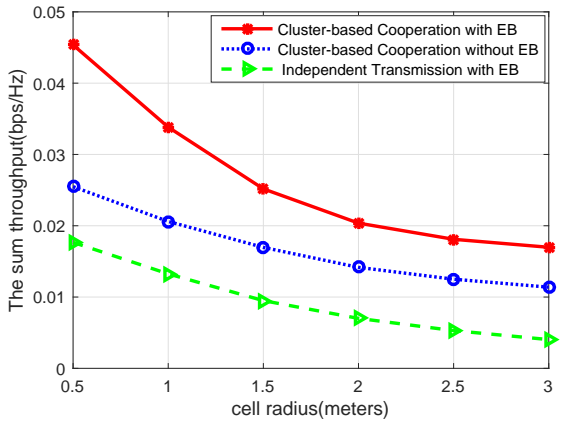}}
\caption{Performance comparison of the different transmission schemes when $d=6$ and the cluster radius $r$ varies. The figures above and below compare the max-min throughput and sum throughput, respectively.}\Large
\label{105} 
\end{figure}

We then compare the throughput performance of the proposed cluster-based cooperation with the two benchmark methods in Section IV.~B. In particular, both max-min throughput (user fairness) and sum throughput (spectral efficiency) are compared. In Fig.~\ref{105}, we first investigate the impact of intra-cluster communication links to the overall throughput performance by fixing $d=6$ and varying $r$. We can see that all the schemes are very sensitive to the degradation of intra-cluster communication links, where both the max-min throughput and sum throughput have dropped by more than $50\%$ for all the schemes when $r$ increases from $1$ to $3$. Nonetheless, we can still observe that the max-min throughput drops more quickly than the sum throughput as $r$ increases, because the max-min throughput is directly determined by the users close to the cluster edge. In both Fig.~\ref{105}(a) and (b), we can see the evident advantage of the proposed method compared to the two benchmark methods, where either cooperation or energy beamforming is absent. On average, the proposed cooperation method achieves around $40\%$ higher max-min throughput than that of cooperation without EB, and over $200\%$ higher max-min throughput than that of the independent transmission method. Moreover, the advantage is even more evident in case of sum throughput performance.

\begin{figure}
\centering
\subfigure[Max-min throughput.]{
\label{figa} 
\includegraphics[width=0.45\textwidth]{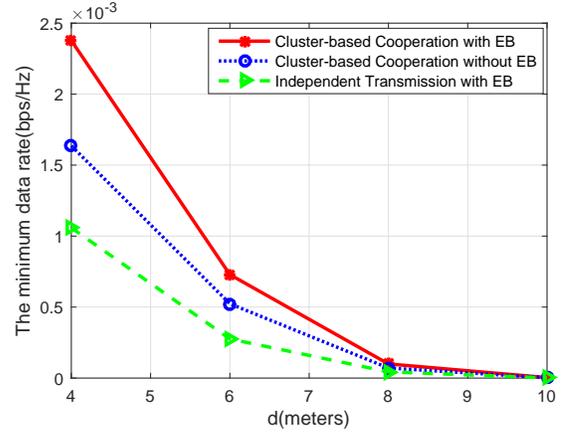}}
\hspace{1in}
\subfigure[Sum throughput.]{
\label{fig:subfig:b} 
\includegraphics[width=0.45\textwidth]{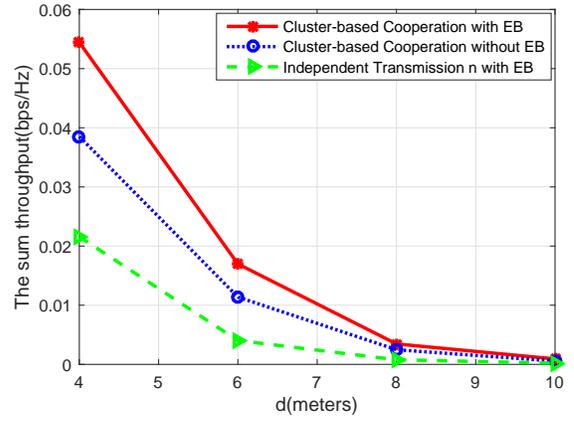}}
\caption{Performance comparison of the different transmission schemes when $r=3$ and the cluster-to-HAP distance $d$ varies. The figures above and below compare the max-min throughput and sum throughput, respectively.}\Large
\label{106} 
\end{figure}

We also investigate in Fig.~\ref{106} the impact of cluster-to-HAP communication links to the overall throughput performance by fixing $r=3$ and varying $d$. Similar to Fig.~\ref{105}, we can see that the proposed cooperation method achieves evident performance advantages over the two benchmark method, especially when the cluster-to-HAP distance is small to moderate, e.g., $d<8$ meters. However, as we further increase $d$, all the schemes achieve very low data rates because of the dramatic energy signal attenuation over distance. The results show that the effective operating range of the considered cooperation method is fundamentally limited by the relatively low efficiency of energy transmissions. In fact, wireless powered communication is \emph{effective only when the power transmission distance is not too large}, such that the WDs can harvest sufficient energy to perform information transmission. In practice, we can improve the performance by several methods, e.g., increasing the number of antennas of the HAP, optimizing the route of a mobile HAP, or increasing the HAP's transmit power. Due to the scope of this paper, we omit the simulations on these performance improving methods. The results in Fig.~\ref{105} and \ref{106} show that the proposed cooperation method can \emph{effectively enhance user fairness and spectral efficiency.}

\begin{figure}
\centering
\subfigure[Max-min throughput.]{
\label{figa} 
\includegraphics[width=0.45\textwidth]{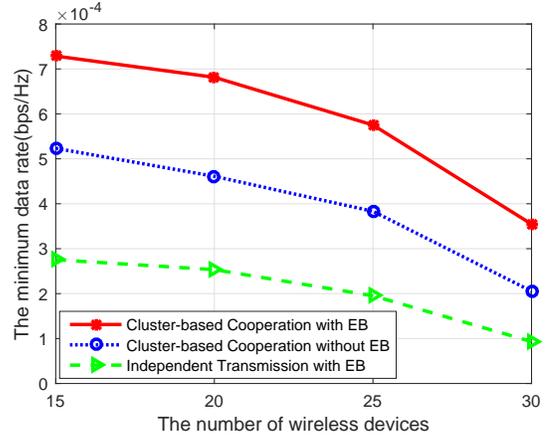}}
\hspace{1in}
\subfigure[Sum throughput.]{
\label{fig:subfig:b} 
\includegraphics[width=0.45\textwidth]{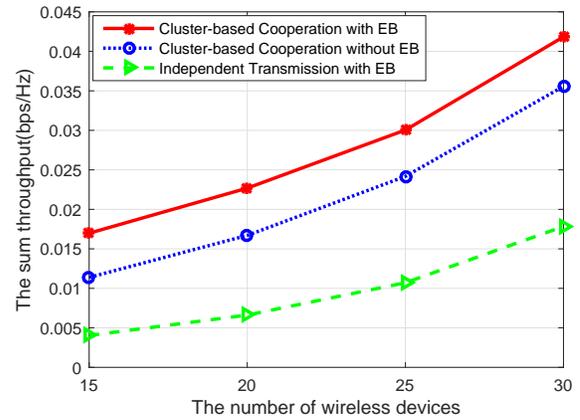}}
\caption{Performance comparison of the different transmission schemes when the number of WDs $N$ varies. The figures above and below compare the max-min throughput and sum throughput, respectively.}\Large
\label{107} 
\end{figure}

In Fig.~\ref{107}, we evaluate the stability of throughput performance when the number of WDs $N$ increases from $15$ to $30$. Without loss of generality, we set $d=6$ and $r=3$. We can see from Fig.~\ref{107}(a) that the max-min throughput decreases with the number of WDs for all the schemes. This is because on average each WD is allocated with shorter transmission time, and thus the data rate of the worst-performing WD decreases. In particular, the decrease of max-min throughput is moderate when $N$ increases from $15$ to $25$, but becoming significant as $N$ further increases. However, we observe in Fig.~\ref{107}(b) that the sum-throughput increases with $N$, although the data rate of each individual may decrease. This indicates that a tradeoff exists between each individual user's throughput and the aggregate network throughput. In practice, \emph{the number of WDs should be kept moderate}, e.g., less than $25$ in the considered network setup. Nonetheless, we can still observe significant performance gain of the proposed method over the two benchmark methods, where the worst-performing WD can still maintain relatively high data rate when the network size is large (e.g., $N=30$).

\vspace{-2ex}

\section{Conclusions}
In this paper, we have proposed a cluster-based cooperation method in a WPCN where a WD is designated as the CH to assist the transmission of other WDs. In particular, energy beamforming technique is applied at the multi-antenna HAP to achieve directional energy transfer to balance the different energy consumption rates of the WDs, especially the high power consumption of the CH. We proposed an efficient algorithm to achieve the optimal max-min throughput among the WDs, by jointly optimizing the EB design, the transmit time allocation among the HAP and the WDs, and the transmit power allocation of the CH. Extensive simulations under practical network setups showed that the proposed method can significantly improve both the user fairness and spectral efficiency compared to non-trivial benchmark methods. Moreover, we also found that the proposed cooperation is most effective when selecting the WD closest to the cluster center as the CH, both the intra-cluster and cluster-to-HAP communication links are strong, and the number of cooperating WDs is moderate.

\end{document}